\documentclass[aps,pre,showpacs,groupedaddress]{revtex4}
\usepackage{amsfonts}
\usepackage{mathrsfs}
\usepackage{dcolumn}
\usepackage{graphicx}
\usepackage{amsmath}
\usepackage{amssymb}
\usepackage{amsbsy}

\def\pcf{{\rm PopC}_{1}}
\def\pcs{{\rm PopC}_{2}}

\begin{document}

\title{Theory of population coupling and applications to describe high order correlations in
large populations of interacting neurons}

\author{Haiping Huang}
\affiliation{RIKEN Brain Science Institute, Wako-shi, Saitama
351-0198, Japan}
\date{\today}

\begin{abstract}
To understand the collective spiking activity in neuronal populations, it is essential to reveal basic
circuit variables responsible for these emergent functional states. Here, I develop a mean field theory for the
population coupling recently proposed in the studies of visual cortex of mouse and monkey, relating the individual neuron 
activity to the population activity, and extend the original form to
the second order, relating neuron-pair's activity to the population activity, to explain the high order correlations observed in the neural data. I test the computational framework on
the salamander retinal data and the cortical spiking data of behaving rats. For the retinal data, the original form of population coupling and its advanced form can 
explain a significant fraction of two-cell correlations and three-cell correlations, respectively. For the cortical data, the performance becomes much better, and the 
second order population coupling reveals non-local effects in local cortical circuits.
\end{abstract}

 \maketitle

\section{Introduction}
To uncover the neural circuit mechanisms underlying animal behavior, e.g., working memory or
decision making, is a fundamental issue in systems neuroscience~\cite{Quiroga-2009,Yuste-2015}. Recent developments in multi-neuron recording
methods make simultaneous recording of neuronal population activity possible, which gives rise to the
challenging computational tasks of finding basic circuit variables responsible for the observed 
collective behavior of neural populations~\cite{Stev-2011}. The collective behavior arises from interactions among neurons, and forms the high dimensional neural code. To search 
for a low dimensional and yet neurobiologically plausible representation of the neural code, thus becomes a key step
to understand how the collective states generate behavior and cognition.  

Correlations among neurons' spiking activities play a prominent role in deciphering the neural code~\cite{Cohen-2011}.
Various models were proposed to understand the pairwise correlations in the population activity~\cite{Bialek-2006,Cocco-09,Opper-2011}.
Modeling these correlations sheds light on the functional organization of the nervous system~\cite{Ganmor-2011jns}. 
However, as the population size grows, higher order correlations have to be taken into account for
modeling synchronous spiking events, which are believed to be crucial for neural 
information transmission~\cite{Martignon-2000,Markus-2003,Victor-2010}. In addition, the conclusion drawn from small size populations
may not be correct for large size populations. Theoretical studies have already proved that
high order interactions among neurons are necessary for generating widespread population activity~\cite{Amari-2003,Montani-2013}.
However, introduction of high order multi-neuron couplings always suffers from a combinatorial explosion of model
parameters to be estimated from the finite neural spike train data. 

To account for high order correlations,
various models with different levels of approximation were proposed, for example, the reliable interaction model~\cite{Ganmor-2011} with
the main caveat that the rare patterns are discarded during inference of the coupling terms, the dichotomized
Gaussian model~\cite{Opper-2011,Yu-2011} in which correlations among neurons are caused by common Gaussian 
inputs to threshold neurons, the K-pairwise model~\cite{Tkacik-2013,Tkacik-2014} in which an effective
potential related to the synchronous firing of $K$ neurons was introduced, yet hard to be interpreted in terms of 
functional connectivity, and the restricted Boltzmann machine~\cite{Koster-2014} where hidden units
were shown to be capable of capturing high order dependences but their number should be predefined and difficult
to infer from the data~\cite{Huang-2015JPA}. One can also take into account part of the statistical features of the population
activity (e.g., simultaneous silent neural pattern) and assume homogeneity for high
order interactions among neurons due to the population size limitation~\cite{Shimazaki-2015}. In this paper, I provide a low
dimensional neurobiological model for describing the high order correlations and extracting useful information
about neural functional organization and population coding.

In this study, I interpret correlations in terms of population coupling, a concept recently proposed to
understand the multi-neuron firing patterns of the visual cortex of mouse and monkey~\cite{Okun-2015}.
The population coupling characterizes the relationship of the activity of a single neuron with the population activity;
this is because, the firing of one neuron is usually correlated with the firing pattern of other neurons. I further generalize 
the original population coupling to its higher order form, i.e., the relationship of pairwise firing with the population activity. I then 
derive the practical dimensionality reduction method for both types of population couplings, and test the method on different types of neural data, including ganglion cells in the salamander retina onto which a repeated
natural movie was projected~\cite{Tkacik-2014}, and layer 2/3 as well as layer 5 cortical cells in the medial prefrontal cortex (MPC) of behaving rats~\cite{Fuji-2008}.	

In this paper, I develop a theoretical model of population coupling and its advanced form,
to explain higher order correlations in the neural data. Methodologically, 
I propose the fast mean field method not only to learn the population couplings but also to evaluate the high order correlations. 
Note that this is computationally hard in a traditional maximum entropy model by using sampling-based method. Conceptually, 
I generalize the normal population coupling by introducing the second-order population coupling, 
which reveals interesting features from the data. First, it can explain a significant amount of three-cell correlations,
and it works much better in cortical data than in retinal data. Second, 
the second-order population coupling matrix has distinct features in retinal and cortical data.
The cortical one shows clear stripe-like structure while the retinal one has no such apparent structure. 
Altogether, this work marks a major step to understand the low-order representation of complex neural activity in both concepts and methods.
\section{Model description and mean-field methods}
\label{model}
For a neuronal population of size $N$, the neural spike trains of duration $T$ are binned with temporal resolution $\tau$, yielding $M=\lceil T/\tau\rceil$ samples of
$N$-dimensional binary neural firing patterns. I use $s_i=+1$ to denote firing state of neuron $i$, and $s_i=-1$ for silent state.
Neural responses to repeated stimulus (or the same behavioral tasks) vary substantially (so-called trial-to-trial variability)~\cite{Bialek-1997,Okun-2012}.
To model the firing pattern statistics, I assign each firing pattern $\mathbf{s}$ a cost function (energy in statistical physics jargon) $E(\mathbf{s})$, then 
the probability of observing one pattern $\mathbf{s}$ can be written as $P(\mathbf{s})\propto\exp(-E(\mathbf{s}))$, where
\begin{equation}\label{energy01}
E(\mathbf{s})=-\sum_{i}h_is_i-\sum_{i}J_is_i\Biggl(\sum_{j\neq i}s_j\Biggr).
\end{equation}
This is the first low dimensional representation to be studied. High
energy state $\mathbf{s}$ corresponds to low probability of observation. $h_i$ is the firing bias constraining the firing rate of neuron $i$, while
$J_i$ characterizes how strongly neuron $i$'s spiking activity correlates with the population activity measured by the sum of other neurons' activity.
I name $J_i$ the first order population coupling ($\pcf$). Thus, only $2N$ parameters needs to be estimated from the neural data. This number of 
model parameters is much less than that in conventional maximum entropy model~\cite{Tkacik-2014}.

To model the high order correlation (e.g., three neuron firing correlation), I further generalize $\pcf$ to its advanced form, i.e., the second
order population coupling, namely $\pcs$, describing the relationship of pairwise firing with the population activity, and the corresponding energy
is given by
\begin{equation}\label{energy02}
E(\mathbf{s})=-\sum_{i}h_is_i-\sum_{i<j}w_{ij}s_is_j\Biggl(\sum_{k\neq i,j}s_k\Biggr),
\end{equation}
where $w_{ij}$ characterizes how strongly the firing state of the neuron pair $(ij)$ correlates with the firing activities of other neurons.
This term is expected to increase the prediction ability for modeling high order correlations in the neural data. Under the framework of
$\pcs$, the total number of parameters to be estimated from the data is $N+N(N-1)/2$. $\pcf$ and $\pcs$ have a clear neurobiological interpretation (for $\pcf$,
see a recent study~\cite{Okun-2015}, and the results obtained under the $\pcs$ can also be experimentally tested), and moreover they
can be interpreted in terms of functional interactions among neurons (as shown later).

To find the model parameters as a low dimensional representation, I apply the maximum likelihood learning principle corresponding to maximizing
the log-likelihood $\ln P(\mathbf{s})$ with respect to the parameters. The learning equation for $\pcf$ is given by
\begin{subequations}\label{LE01}
\begin{align}
h_i^{t+1}&=h_i^{t}+\eta\Biggl(\Bigl<s_i\Bigr>_{{\rm data}}-\Bigl<s_i\Bigr>_{{\rm model}}\Biggr),\\
J_i^{t+1}&=J_i^{t}+\eta\Biggl(\sum_{j\neq i}\Bigl<s_is_j\Bigr>_{{\rm data}}-\sum_{j\neq i}\Bigl<s_is_j\Bigr>_{{\rm model}}\Biggr),
\end{align}
\end{subequations}
where $t$ and $\eta$ denote the learning step and learning rate, respectively. The maximum likelihood learning shown here has a simple 
interpretation of minimizing the Kullback-Leibler divergence between the observation probability and the model probability~\cite{Cocco-2012,Huang-2013epjb}.
In an analogous way, one gets the learning equation for $\pcs$,
\begin{equation}\label{LE02}
w_{ij}^{t+1}=w_{ij}^{t}+\eta\Biggl(\sum_{k\neq i,j}\Bigl<s_is_js_k\Bigr>_{{\rm data}}-\sum_{k\neq i,j}\Bigl<s_is_js_k\Bigr>_{{\rm model}}\Biggr).
\end{equation}
In the learning equations Eq.~(\ref{LE01}) and Eq.~(\ref{LE02}), the data dependent terms can be easily computed from the binned neural data.
However, the model expectations of the firing rate (magnetization in statistical physics) and correlations are quite hard to evaluate without
any approximations. Here I use the mean field method to tackle this difficulty. 

First, I write the energy term into a unified form,
\begin{equation}\label{energyIsing}
E(\mathbf{s})=-\sum_{i}h_is_i-\sum_{a}\Gamma_a\prod_{i\in\partial a}s_i,
\end{equation}
where $a$ denotes the interaction index and $\partial a$ denotes the neuron set involved in the interaction $a$. $a=(ij)$ for $\pcf$ and
$(ijk)$ for $\pcs$. Therefore, $\pcf$ introduces the pairwise interaction as $\Gamma_{ij}=J_i+J_j$, while $\pcs$ introduces the triplet
interaction as $\Gamma_{ijk}=w_{ij}+w_{jk}+w_{ik}$. The multi-neuron interaction in the conventional Ising model is decomposed into
first order or second order population coupling terms. This decomposition still maintains the functional heterogeneity of single neurons or neuron pairs, but reduces
drastically the dimensionality of the neural representation for explaining high order correlations. In principle, one can combine $\pcf$ and $\pcs$ to predict 
both pairwise and triplet correlations. However, in this work, I focus on the pure effect of each type of
population coupling.

In fact, the conventional Ising model~\cite{Roudi-2009} can be recovered by setting $\Gamma_{ij}=J_{ij}$, which is pairwise interaction. The learning equation is derived
similarly, and is run by reducing the deviation between the model pairwise correlation and the clamped one (computed from the data)~\cite{Huang-2016}.
 
Second, the statistical properties of the model (Eq.~\ref{energyIsing}) can be analyzed by the cavity method in the mean field theory~\cite{cavity-2001}.
The self-consistent equations are written in the form of message passing (detailed derivations were given in Refs~\cite{Huang-2009pre,MM-2009}, see also Appendix~\ref{app:Deriv}) as
\begin{subequations}\label{bp}
\begin{align}
m_{i\rightarrow a}&=\tanh\left(h_i+\sum_{b\in\partial i\backslash a}\tanh^{-1}\hat{m}_{b\rightarrow i}\right),\\
\hat{m}_{b\rightarrow i}&=\tanh\Gamma_b\prod_{j\in\partial b\backslash i}m_{j\rightarrow b},
\end{align}
\end{subequations}
where $\partial b\backslash i$ denotes the member of interaction $b$ except $i$, and $\partial i\backslash a$ denotes the interaction set
$i$ is involved in with $a$ removed. $m_{i\rightarrow a}$ is interpreted as the message passing from the neuron $i$ to the interaction $a$ it
participates in, while $\hat{m}_{b\rightarrow i}$ is interpreted as the message passing from the interaction $b$ to its member $i$. This iteration equation
is also called the belief propagation (BP), which serves as the message passing algorithm for the statistical inference of the model parameters. Iteration of 
the message passing equation on the inferred model would converge to a fixed point corresponding to a global minimum of the free
energy (in the cavity method approximation~\cite{MM-2009})
\begin{equation}\label{freeE}
F\equiv-\ln Z=-\sum_{i}\ln Z_i+\sum_{a}(|\partial a|-1)\ln Z_a,
\end{equation}
where $Z$ is the normalization constant of the model probability $P(\mathbf{s})$. The free energy contribution of one neuron is 
$-\ln Z_i=-\ln\sum_{x=\pm1}\mathcal{H}_i(x)$ and the free energy contribution of one interaction is $-\ln Z_a=-\ln\cosh\Gamma_a-\ln\left(1+\tanh\Gamma_a\prod_{i\in\partial a}m_{i\rightarrow a}\right)$.
I define the function $\mathcal{H}_i(x)\equiv e^{xh_i}\prod_{b\in\partial i}\cosh\Gamma_b(1+x\hat{m}_{b\rightarrow i})$. At the same time, the model firing rate and multi-neuron
correlation can be estimated as
\begin{subequations}\label{magcorre}
\begin{align}
m_{i}&=\tanh\left(h_i+\sum_{b\in\partial i}\tanh^{-1}\hat{m}_{b\rightarrow i}\right),\\
C_a&=\frac{\tanh\Gamma_a+\prod_{i\in\partial a}m_{i\rightarrow a}}{1+\tanh\Gamma_a\prod_{i\in\partial a}m_{i\rightarrow a}}.
\end{align}
\end{subequations}
Magnetization and correlation are defined as $m_i=\left<s_i\right>$ and $C_a=\left<\prod_{i\in\partial a}s_i\right>$, respectively.

A brief derivation of Eq.~(\ref{magcorre})
is given in Appendix~\ref{app:Deriv}. Here the multi-neuron correlation is calculated directly from the cavity method approximation and expected to be accurate enough for current 
neural data analysis~\cite{Huang-2016}. This is because, correlations under the model are evaluated taking into account nearest-neighbor interactions, rather than naive full
independence among neurons. This approximation is expected to work well in a weakly-correlated neural population~\cite{Bialek-2006,Ganmor-2011jns}, where long-range
strong correlations do not develop. A similar application of this principle revealed a non-trivial geometrical structure of population codes in salamander retina~\cite{Huang-2016}. 
Another advantage is the low computation cost. Both the free energy and the pairwise correlations can 
be estimated by the time complexity of the order $\mathcal{O}(N^2)$ for $\pcf$, and $\mathcal{O}(N^3)$ for triplet correlations in $\pcs$, which is one order of magnitude lower than
the tractable model of $\pcf$ recently proposed in Ref.~\cite{Mora-2016}. A more accurate expression could be derived from linear response theory~\cite{Huang-2012pre}
with much more expensive computational cost (increased by an order of magnitude ($\mathcal{O}(N)$).

To estimate the information carried by a neural population, one needs to compute the entropy, which is defined as
$S=-\sum_{\mathbf{s}}P(\mathbf{s})\ln P(\mathbf{s})$, and it measures the capacity of the neural population for information transmission.
The more obvious variability the neural responses have, the larger the entropy value is. The entropy of the model can be estimated from the fixed point of the message passing equation. Based on the standard thermodynamic relation, $S=-F+E$, where
$E$ is the energy of the neural population and given by
\begin{subequations}\label{Energ}
\begin{align}
E&=-\sum_{i}\Delta E_i+\sum_{a}(|\partial a|-1)\Delta E_a,\\
\Delta E_i&=\frac{h_i\sum_{x=\pm1}x\mathcal{H}_i(x)+\sum_{x=\pm1}\mathcal{G}_i(x)}{\sum_{x=\pm1}\mathcal{H}_i(x)},\\
\Delta E_a&=\Gamma_a\frac{\tanh\Gamma_a+\prod_{i\in\partial a}m_{i\rightarrow a}}{1+\tanh\Gamma_a\prod_{i\in\partial a}m_{i\rightarrow a}},\\
\begin{split}
\mathcal{G}_i(x)&=\sum_{b\in\partial i}e^{xh_i}\left[\Gamma_b\sinh\Gamma_b(1+x\hat{m}_{b\rightarrow i})+x\Gamma_b\cosh\Gamma_b(1-\tanh^{2}\Gamma_b)\prod_{j\in\partial b\backslash i}m_{j\rightarrow b}
\right]\\
&\times\prod_{a\in\partial i\backslash b}\cosh\Gamma_a(1+x\hat{m}_{a\rightarrow i}).
\end{split}
\end{align}
\end{subequations}

The basic procedure to infer population couplings is given as follows. At the beginning, all model parameters are assigned zero value. It is followed by three steps: ($i$) Messages are initialized randomly and uniformly in the interval $(-1,1)$.
($ii$) Eq.~(\ref{bp}) are then run until converged, and the magnetizations as well as multi-neuron correlations are estimated using Eq.~(\ref{magcorre}). ($iii$) The estimated magnetizations and correlations 
are used at each gradient ascent learning step (Eq.~(\ref{LE01}) or
Eq.~(\ref{LE02})). When one gradient learning step is finished, another step starts by repeating the above procedure (from ($i$) to ($iii$)).
To learn the higher order population coupling, the damping technique is used to avoid oscillation behavior, i.e., $w_{{\rm new}}=\gamma w_{{\rm new}}
+(1-\gamma)w_{{\rm old}}$ where $\gamma$ is the damping factor taking a small value.

The inferred model can also be used to generate the distribution of spike synchrony, i.e., the probability of $K$ simultaneous spikes. This distribution can
be estimated by using Monte Carlo (MC) simulation on the model. The standard procedure goes as follows. The simulation starts from a random initial configuration $\mathbf{s}_0$, and tries to search for the 
low energy state, then the energy is lowered by a series of elementary updates, and for each elementary update,
$N$ proposed neuronal state flips are carried out. That is, the transition probability from state $\mathbf{s}$ to $\mathbf{s}'$ with only $s_i$ flipped ($s'_{i}=-s_i$)
is expressed as $e^{-2s_iH_i}$ where $H_i=h_i+\sum_{a\in\partial i}\Gamma_a\prod_{j\in\partial a\backslash i}s_j$. The equilibrium samples
are collected after sufficient thermal equilibration. These samples (a total of $20000$ samples in simulations) are finally used to estimate the distribution of spike synchrony.
\begin{figure}
          \includegraphics[bb=0 0 680 508,scale=0.65]{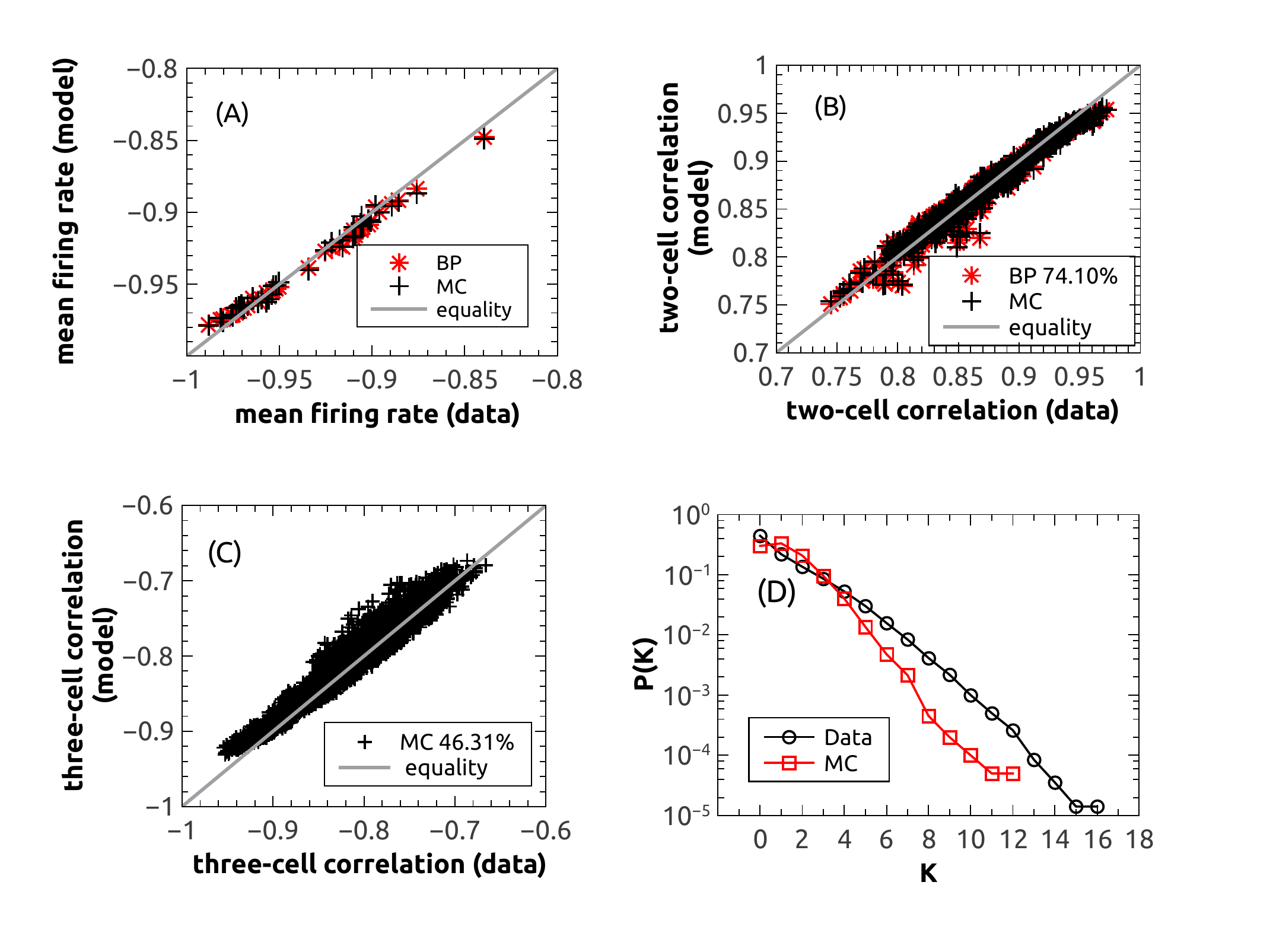}
  \caption{(Color online) Inference performances of $\pcf$ on the retinal data ($N=40$, one typical example). (A) Firing rate in
  the data is reproduced by the model. (B) Two-cell correlations are explained partially by $\pcf$ ($74.10\%$). A Monte-Carlo (MC) 
  sampling of the model yields similar results to belief propagation (BP), which is much faster. (C) From the MC samples,
  three-cell correlations can also be estimated. (D) Probability of $K$ synchronous spiking under the model is compared with that of the data.
     }\label{retinaPC1}
 \end{figure}

\section{Results}
\label{Results}
By using the mean field method, I first test both types of population couplings on the retina data, which is the spike train of $160$ ganglion cells in a small patch of
the salamander retina~\cite{Tkacik-2014}. The retina was stimulated with a repeated natural movie. The spike train data
is binned with the bin size equal to $20 ms$ reflecting the temporal correlation time scale, yielding about $282744$ binary firing
patterns for data modeling.

I then test the same concepts on the cortical data of behaving rats. Rats performed the odor-place matching working memory
during one task session, and spiking activities of $117$ cells in both superficial layer $2/3$ and deep layer (layer $5$) of medial prefrontal cortex were simultaneously recorded (for detailed experiments,
see Ref.~\cite{Fuji-2008}). One task session consists of about $40$ trials, yielding a spike train of these cortical cells binned 
with the temporal resolution $\tau=20 ms$ (a total of $140596$ firing patterns).

\subsection{Inference performances on the retinal data}
Fig.~\ref{retinaPC1} reports the inference result on a network example of $40$ neurons selected randomly from the original 
dataset. The firing rate is predicted faithfully by the model using either MC or BP (Fig.~\ref{retinaPC1} (A)). Inferring 
only $\pcf$, one could predict about $74.10\%$ of entire pairwise correlation (a precision criterion is set to $7.0\times 10^{-3}$ in this paper)
(Fig.~\ref{retinaPC1} (B)). This means that $74.10\%$ of the whole correlation set have the absolute value of the deviation between the predicted correlation
and measured one ($|C_a^{{\rm pred}}-C_a^{{\rm ms}}|$) smaller than the precision criterion. Using the sampled configurations of neural firing activity from the MC simulation, one could also
predict three-cell correlations (Fig.~\ref{retinaPC1} (C)), whereas, the prediction fraction can be improved by a significant amount
after introducing $\pcs$, as I shall show later. In addition, fitting only $2N$ model parameters in $\pcf$ analysis could not predict
the tail of spike synchrony distribution (Fig.~\ref{retinaPC1} (D)); this is expected as no higher order interaction terms are
included in the model, and rare events of large $K$ spikes are also difficult to observe in a finite sampling during MC simulations.

\begin{figure}
          \includegraphics[bb=0 0 758 523, scale=0.6]{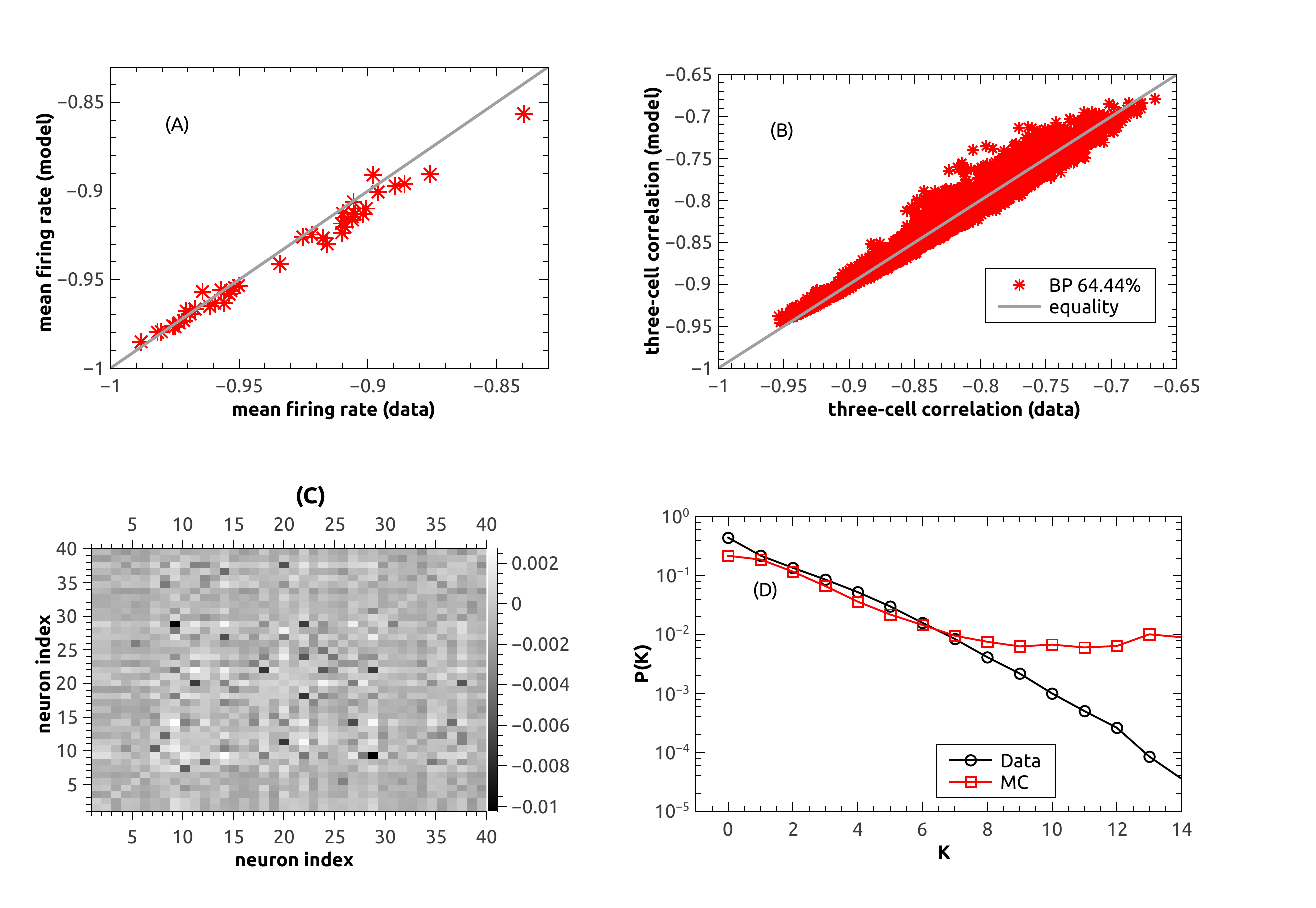}
  \caption{(Color online) Inference performances of $\pcs$ on the retinal data ($N=40$, one typical example). (A) Firing rate in
  the data is reproduced by the model. (B) Three-cell correlations are explained partially by $\pcs$ ($64.44\%$). (C) Interaction
  matrix for $\pcs$. (D) Probability of $K$ synchronous spiking under the model is compared with that of the data.
     }\label{retinaPC2}
 \end{figure}

The inference results of $\pcs$ are given in Fig.~\ref{retinaPC2}. Note that, by considering the correlation between the 
pairwise firing activity and the global population activity, i.e., the second order population coupling, the three-cell
correlation could be predicted partially ($64.44\%$), and this fraction is much larger than that of $\pcf$ (Fig.~\ref{retinaPC1} (C)). This is due to
the specific structure of $\pcs$, which incorporates explicitly three-cell correlations into the construction of couplings (Eq.~(\ref{LE02})). Technically, the mean-field theory for
$\pcs$ avoids the slow sampling and evaluates the high order correlations in a fast way. Alternatively, one could fit the data using the conventional Ising
model~\cite{Roudi-2009} with the same number of model parameters as $\pcs$, whereas, the three-cell correlations are hard to predict using MC samplings, and a similar phenomenon was
also observed in a previous work for modeling pairwise correlations~\cite{Huang-2012pre}. Therefore I speculate that $\pcs$ 
acts as a key circuit variable for third order correlations. 

\begin{figure}
          \includegraphics[bb=0 0 659 244, scale=0.6]{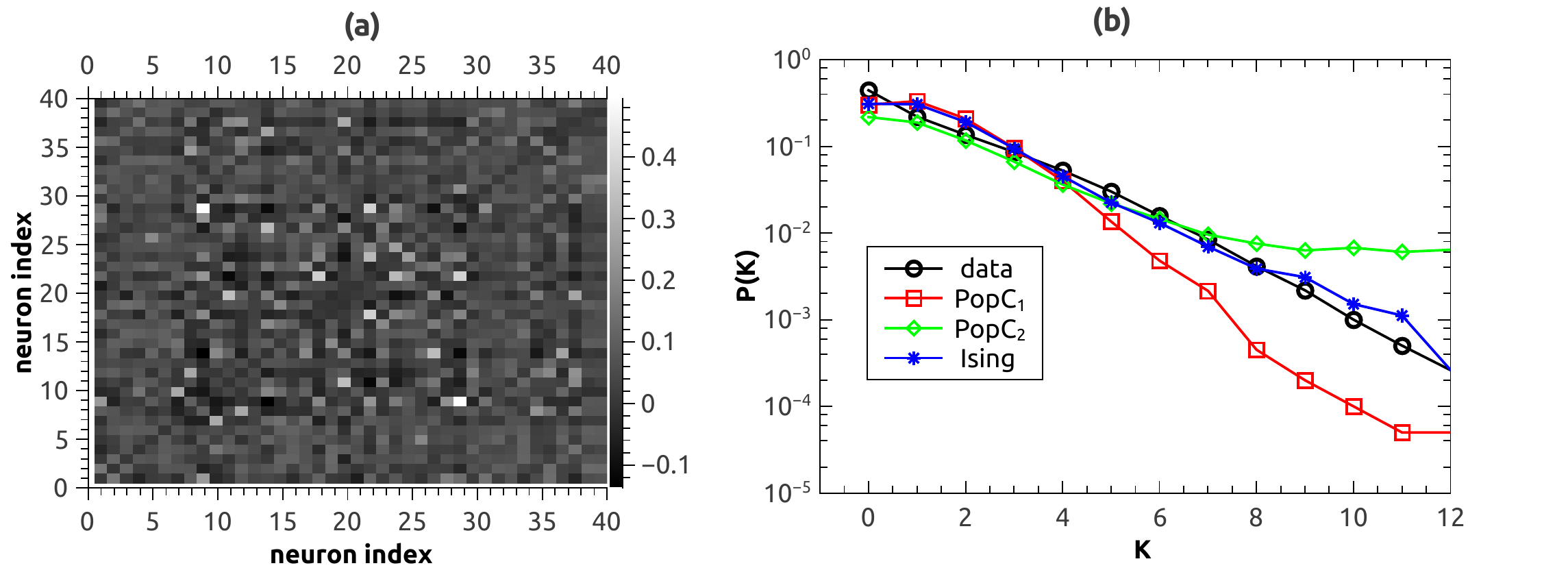}
  \caption{(Color online) Inference performances of Ising model on the retinal data ($N=40$, one typical example) compared with population coupling. (a) Two-neuron interaction
  matrix. (b) Probability of $K$ synchronous spiking under the model is compared with that of the data.
     }\label{retinaIsing}
 \end{figure}

The interaction matrix of $\{w_{ij}\}$ reveals 
how important each pair of neurons is for the entire population activity (emergent functional state of the whole network).
As shown in Fig.~\ref{retinaPC2} (C), $\pcs$ matrix has no apparent structure of organization, i.e., each neuron can be paired with both positive and negative couplings.
Some pairs have large negative $\pcs$, suggesting that these components are anti-correlated with the population activity. That is to say, the activity of these neuron-pairs is
not synchronized to the population activity characterized by the summed activity over all neurons except these pairs. In the network, there 
also exist positive $\pcs$s, which shows that these neuron-pairs are positively correlated with the population in neural activity. The interaction matrix shown here 
may be related to the revealed overlapping modular structure of retinal neuron interactions~\cite{Ganmor-2011jns,Ganmor-2011}. In this structure,
neurons interact locally with their adjacent neurons, and in particular this feature is scalable and applicable for larger networks. It seems that one individual
neuron does not impact directly the entire population, and a small group of neighboring neurons have similar visual feature selectivity~\cite{hetero-2011}. This result
is also consistent with two-neuron interaction map of the conventional Ising model (Fig.~\ref{retinaIsing} (a)). Note that in functional interpretation, these 
two-neuron interactions are inherently different from $\pcs$, which is designed to explain high-order correlations by using less model parameters than necessary.

$\pcs$ behaves better than $\pcf$ in predicting the spike synchrony distribution (Fig.~\ref{retinaPC2} (D)) in the small $K$ regime (the prediction is improved
from $K=4$ for $\pcf$ to $K=8$ for $\pcs$). An intuitive explanation is that $\pcs$ introduces equivalently triplet interactions among neurons, and
it is known that high order interactions are necessary for generating widespread population activity~\cite{Amari-2003}.  However,
$\pcs$ overestimates the distribution
when rare events of synchronous spiking are considered. This may be related to the difficulty of obtaining sufficient equilibrium samples of the model, especially those samples with
large population activity. The spike synchrony distribution is also compared with that obtained under Ising model (Fig.~\ref{retinaIsing} (b)). Different performances are
related to the multi-information measure of neural population explained below.

\begin{figure}
          \includegraphics[bb=0 0 340 265,scale=0.65]{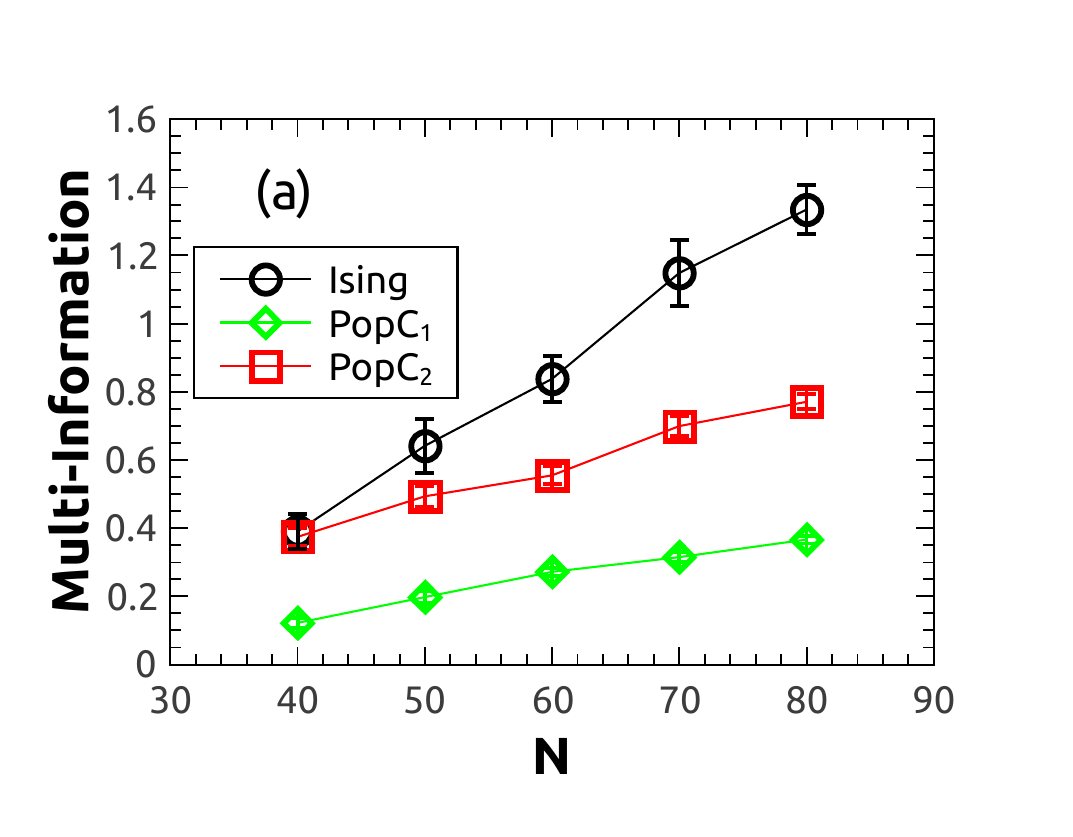}
     \hskip .1cm
     \includegraphics[bb=0 0 351 265,scale=0.65]{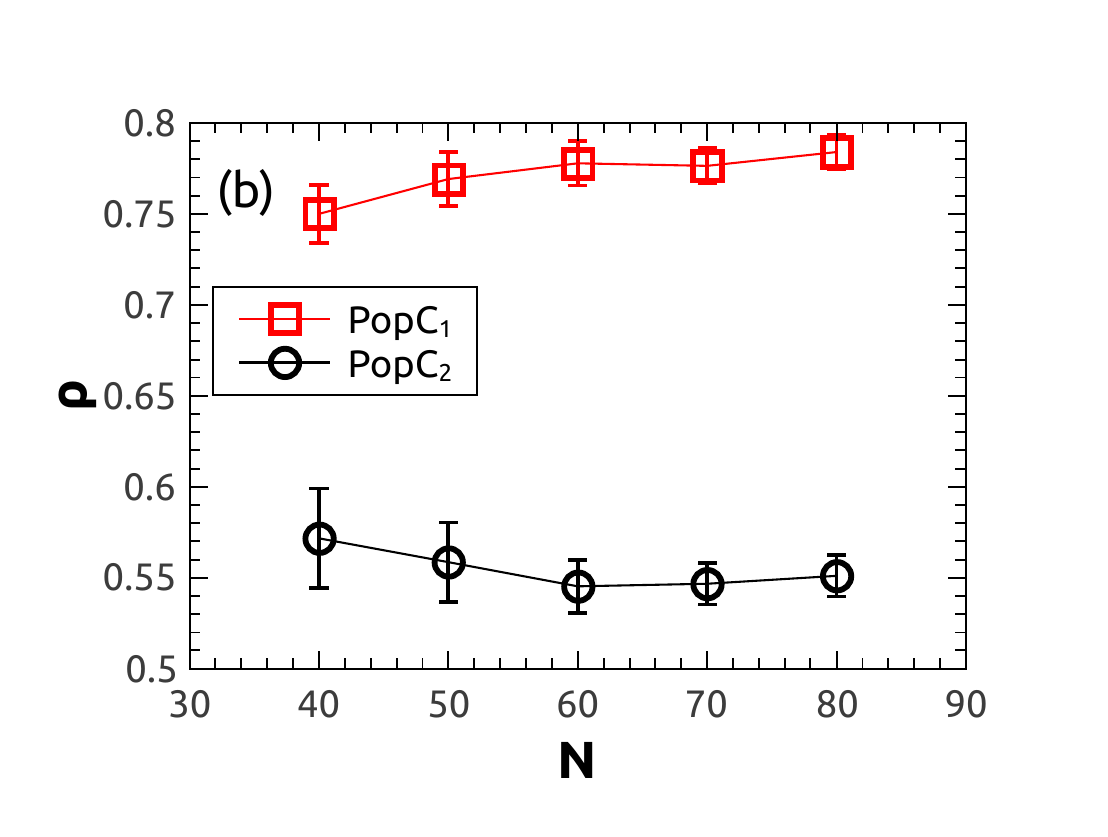}
     \vskip .1cm
  \caption{(Color online) Multi-information and prediction fraction of correlations under the simplified model for the retinal network.
  The result is averaged over $10$ network samples for
  each $N$. (a) Multi-information (in bits) versus the network size $N$.  (b) The prediction fraction $\rho$ versus the network size. $\pcf$ is used to predict two-cell correlations, and
  $\pcs$ is used to predict three-cell correlations.
     }\label{retinaIF}
 \end{figure}

The amount of statistical structure in the neural data due to introducing interactions among neurons can be measured by
the multi-information~\cite{Bialek-2006}. I first introduce an independent model where only the firing rates of individual neurons are fitted and 
the corresponding entropy is defined as $S_{{\rm ind}}$. The multi-information is then defined as $I(N)=S_{{\rm ind}}-S_{{\rm model}}$, in which
$S_{\rm ind}=\sum_i\sum_{x=\pm1}\mathcal{S}((1+m_ix)/2)$, where $\mathcal{S}(u)=-u\ln u$, and $S_{{\rm model}}$ is assumed to be an upper bound to the true entropy. The true entropy for large populations is difficult to
estimate since it requires including all possible interactions among neurons. However, the model entropy with low order interaction
parameters could be an approximate information capacity for the neural population, which depends on how significant the higher order
correlations are in the population. 

Fig.~\ref{retinaIF} (a) shows the multi-information as a function of the network size. $\pcf$ and $\pcs$ are 
compared with the Ising model~\cite{Huang-2012pre}, which reconstructs faithfully the pairwise correlations. $\pcs$ improves
significantly over $\pcf$ in capturing the information content of the network, but its multi-information is still below that of
the Ising model, which is much more evident for larger network size. This is expected, because only part of third order
correlations are captured by $\pcs$, while the Ising model describes accurately the entire pairwise correlation profile which may be the main contributor 
to the collective behavior observed in the population. However, $\pcs$ provides us an easy way to understand the higher order
correlation, while in the Ising model, it is computationally difficult to estimate the higher order correlations. The average 
prediction fraction of correlations by $\pcf$ and $\pcs$ is plotted in Fig.~\ref{retinaIF} (b). $\pcf$ predicts more than $75\%$ of the pairwise correlations,
while $\pcs$ predicts more than $54\%$ of the triplet correlations. The prediction fraction changes slightly with the network size.

\begin{figure}
          \includegraphics[bb=0 0 946 505,scale=0.6]{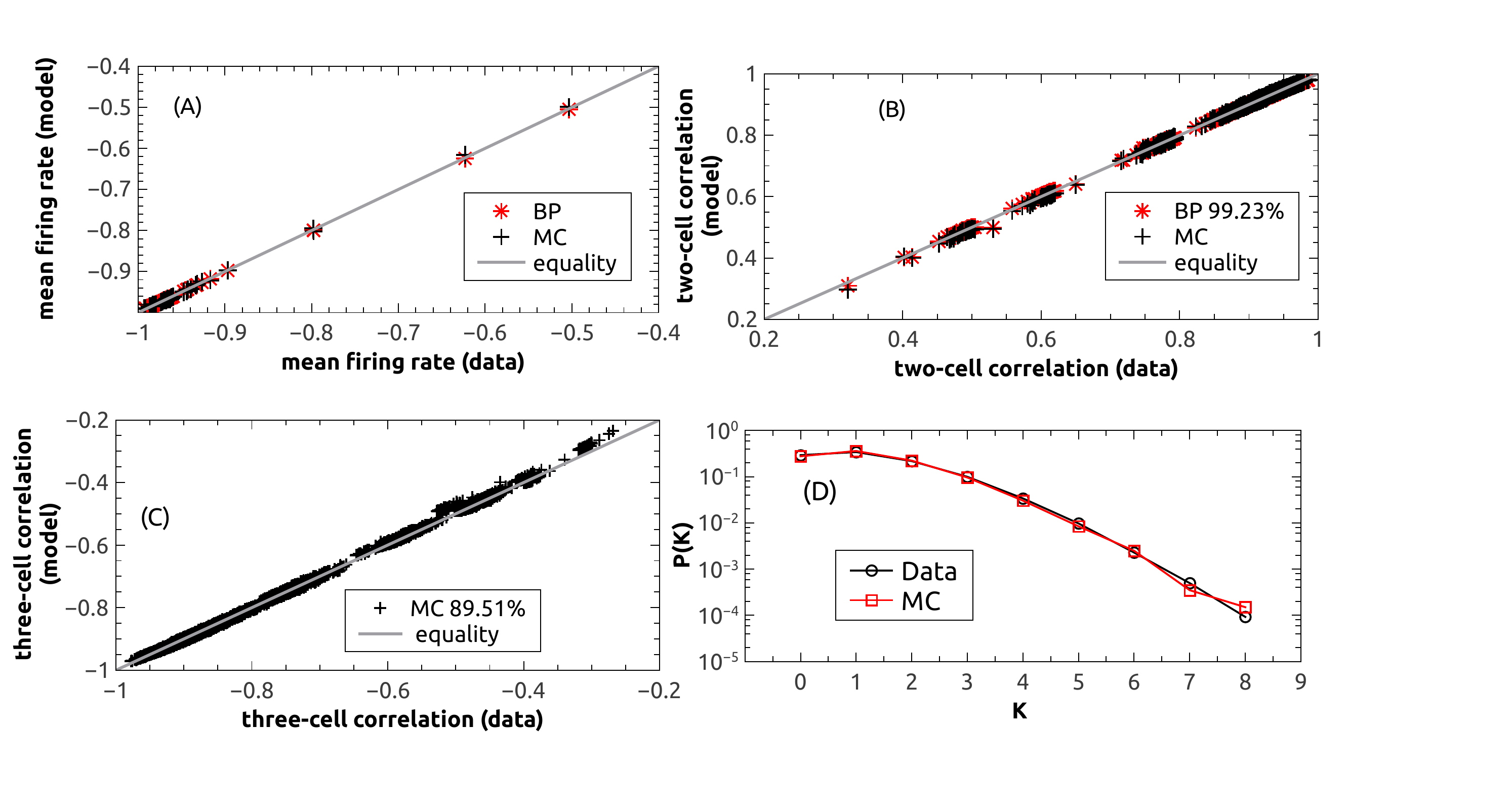}
  \caption{(Color online) Inference performances of $\pcf$ on the cortical data ($N=40$, one typical example). (A) Firing rate in
  the data is reproduced by the model. (B) Two-cell correlations are explained partially by $\pcf$ ($99.23\%$). A Monte-Carlo (MC) 
  sampling of the model yields similar results to belief propagation (BP). (C) From the MC samples,
  three-cell correlations can also be estimated. (D) Probability of $K$ synchronous spiking under the model is compared with that of the data.
     }\label{mpcPC1}
 \end{figure}
 
\begin{figure}
          \includegraphics[bb=0 0 758 523, scale=0.6]{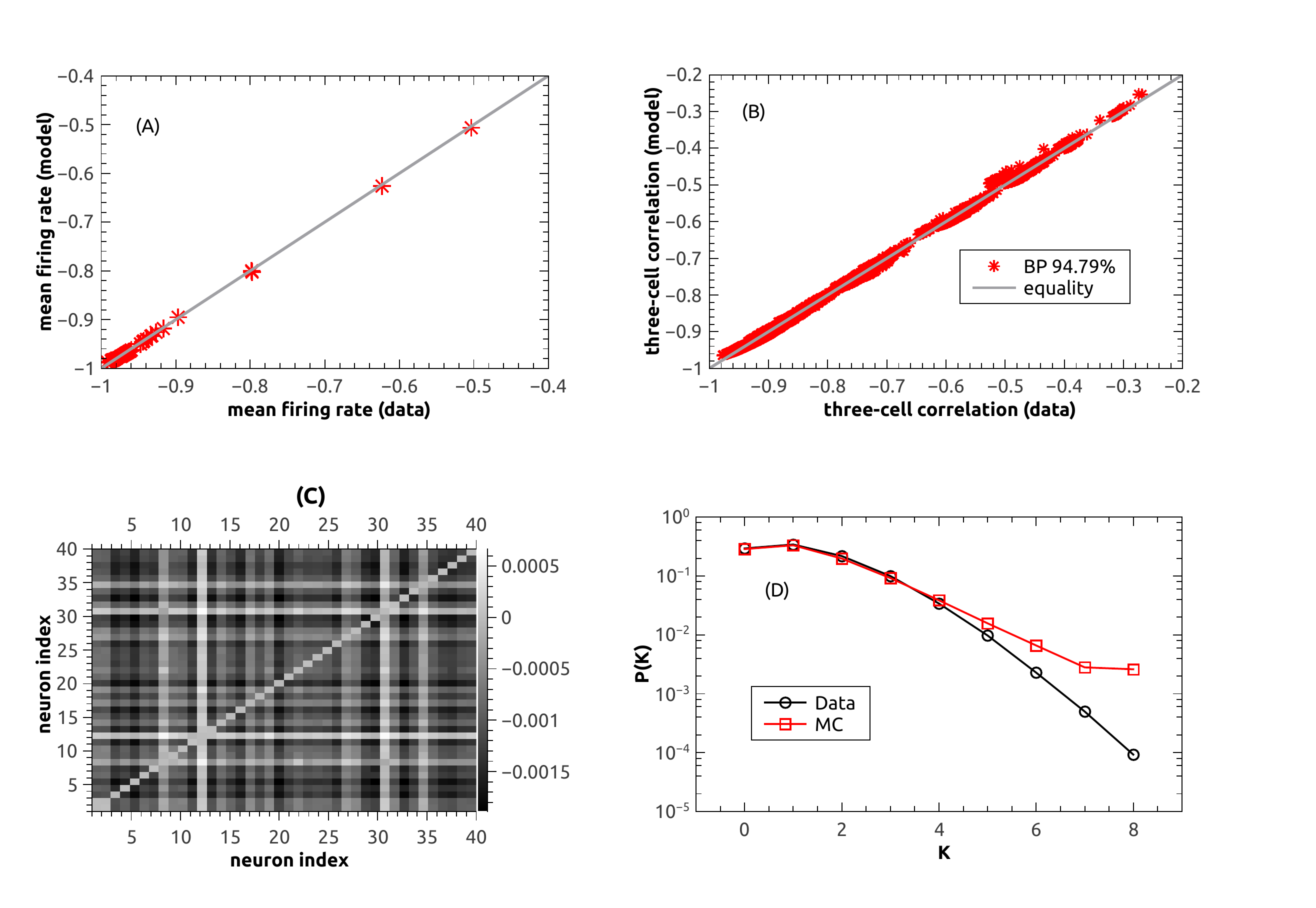}
  \caption{(Color online) Inference performances of $\pcs$ on the cortical data ($N=40$, one typical example). (A) Firing rate in
  the data is reproduced by the model. (B) Three-cell correlations are explained partially by $\pcs$ ($94.79\%$). (C) Interaction
  matrix for $\pcs$. (D) Probability of $K$ synchronous spiking under the model is compared with that of the data.
     }\label{mpcPC2}
 \end{figure}
 
\subsection{Inference performances on the cortical data}

To show the inference performance of both types of population couplings on the cortical data, I randomly select a typical network example of $40$
neurons from the original dataset, and then apply the computation scheme to this typical example. Results are shown in 
Fig.~\ref{mpcPC1}. Surprisingly, the simplified $\pcf$ is able to capture as high as $99.23\%$ of pairwise correlations, implying that
when a rat performed working memory tasks, there exists a simplified model to describe emergent functional states in
the medial prefrontal cortical circuit. Moreover, MC sampling of the $\pcf$ model also predicts well the spike synchrony distribution (Fig.~\ref{mpcPC1} (D)).
This is very different from that observed in the retinal data. In this sense, the MPC circuit is simple in 
its functional states when the subject is performing specified tasks. 

\begin{figure}
          \includegraphics[bb=0 0 659 244, scale=0.6]{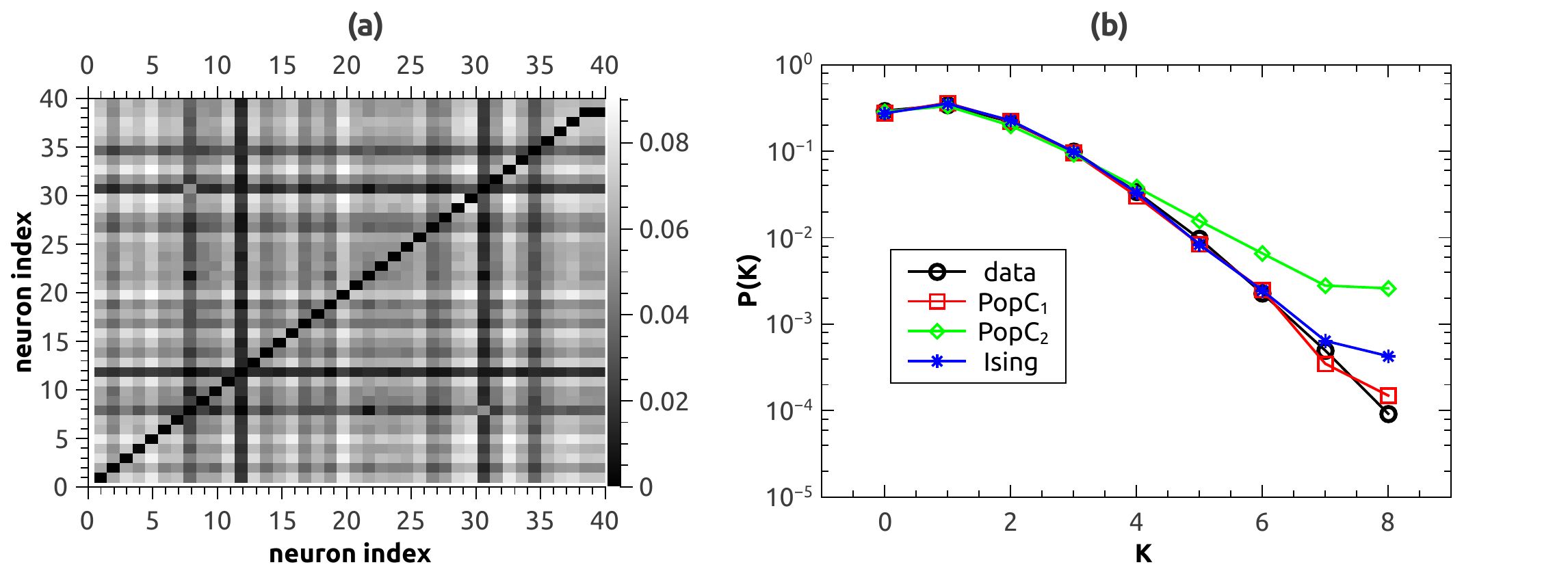}
  \caption{(Color online) Inference performances of Ising model on the cortical data ($N=40$, one typical example) compared with population coupling. (a) Two-neuron interaction
  matrix. (b) Probability of $K$ synchronous spiking under the model is compared with that of the data.
     }\label{cortexIsing}
 \end{figure}

More interesting circuit features are revealed by $\pcs$, which is 
shown in Fig.~\ref{mpcPC2}. About $94.79\%$ of three-cell correlations are explained by $\pcs$ in the MPC circuit. The interaction
matrix of $\pcs$ in Fig.~\ref{mpcPC2} (C) shows a clear non-local structure in the cortical circuit (stripe-like structure). That is, some neurons interact strongly with nearly all the other
neurons in the selected population, and these interactions have nearly identical strength of $\pcs$. Such neurons having stripe-like structure in the $\pcs$ matrix
may receive a large number of excitatory inputs from pyramidal neurons~\cite{Fuji-2008}, and thus play a key role in shaping the collective spiking behavior during
the working memory task. The non-local effects are consistent with findings reported in the original experimental 
paper (cross-correlogram analysis)~\cite{Fuji-2008} and the two-neuron interaction map under Ising model (Fig.~\ref{cortexIsing} (a)). Thus, to some extent, $\pcs$ may reflect intrinsic connectivity in the cortical circuit, although the relationship between
functional
connections and anatomical connections has not yet been well established~\cite{Yat-2015}. Lastly, $\pcs$ overestimates
the tail of the spike synchrony distribution (Fig.~\ref{mpcPC2} (D)), which may be caused by the sampling difficulty of the inferred model (a model with triplet interactions among
its elements). The spike synchrony distribution of Ising model is also compared (Fig.~\ref{cortexIsing} (b)).
 
 \begin{figure}
    \includegraphics[bb=0 0 337 240,scale=0.65]{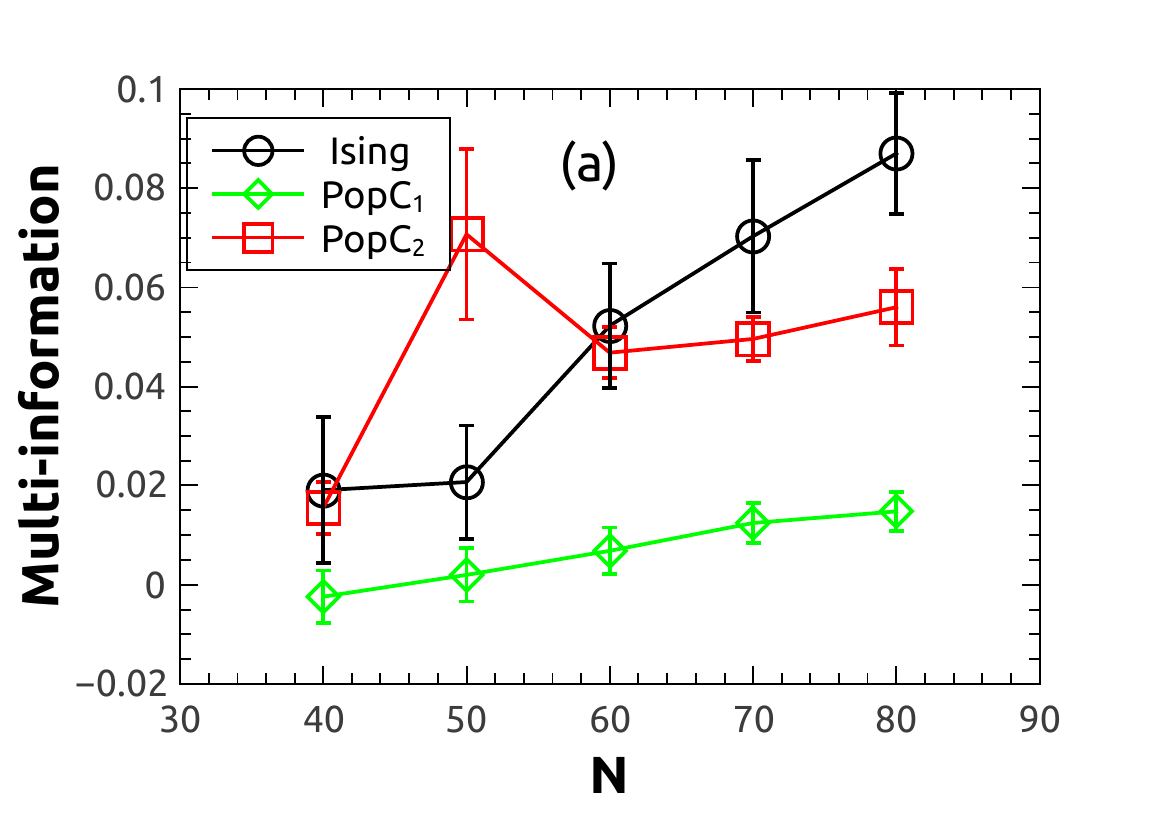}
     \hskip .1cm
  \includegraphics[bb=0 0 320 241,scale=0.7]{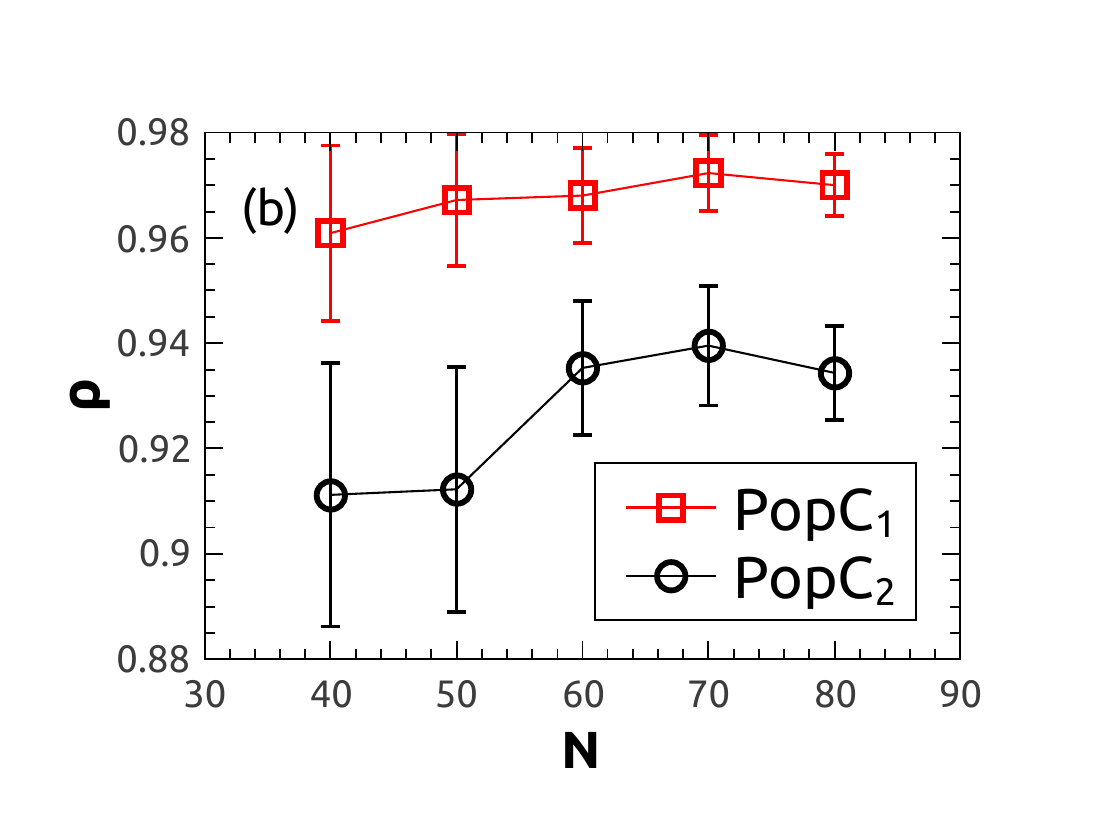}
     \vskip .1cm
  \caption{(Color online) Multi-information and prediction fraction of correlations under the simplified model for the cortical network.
  The result is averaged over $10$ network samples for
  each $N$. (a) Multi-information (in bits) versus the network size $N$. 
   (b) The prediction fraction $\rho$ versus the network size. $\pcf$ is used to predict two-cell correlations, and
  $\pcs$ is used to predict three-cell correlations.
     }\label{mpcIF}
 \end{figure}
 
 Multi-information versus the cortical network size is plotted in Fig.~\ref{mpcIF} (a). In the cortical circuit, $\pcs$ behaves
 comparably with the Ising model; even for some network size ($N=50$), it reports a higher information content than the Ising model in the randomly selected
 subpopulations, which may be caused by the nature of the selected neurons (e.g., inhibitory interneurons~\cite{Fuji-2008}, and they have stripe-like structure in the $\pcs$ matrix).
 Note that $\pcf$ gives an information close to zero for small network sizes, suggesting that
 by introducing $\pcf$, one could not increase significantly the amount of statistical structure in the network activity explained by the model. However,
 the multi-information of $\pcf$ grows with the network size, indicating that the role of $\pcf$ would be significant for larger neural populations.
 Fig.~\ref{mpcIF} (b) reports the prediction fraction of the correlation profile by applying $\pcf$ and $\pcs$. Both population couplings can
 capture over $90\%$ of correlations, which is significantly different from that observed in the retinal data. 
 
 \section{Discussion}
 \label{Disc}
The emergent properties of the neural code arise from interactions among individual neurons. A complete characterization of the population
activity is difficult, because on the one hand, the number of potential interactions suffers from a combinatorial explosion, on 
the other hand, the collective behavior at the network level would become much more complex as the network size grows. In this paper,
I develop a theoretical framework to understand how pairwise or higher order correlations arise and the basic circuit variables 
corresponding to these correlation structures. The model is based on the concept of population coupling, characterizing 
the relationship between local firing activity of individual neuron or neuron-pair and the global neural activity. An advantage is that,
it provides a low dimensional and neurobiologically interpretable representation to understand the functional interaction between 
neurons and their correlation structures. In particular, the concept of population coupling and the associated mean field method used in this paper offer an easy way to evaluate higher order correlations, while
the usual sampling method is computationally hard and traditional models (e.g., Ising model) lack a direct interpretation of higher order correlations in terms of simplified (population) couplings. 

With the mean field method, the concept of population coupling is tested on two different types of neural data. One is the firing neural activities of retinal ganglion cells under
natural movie stimuli. The other is the population activities of medial prefrontal cortex when a rat was performing odor-place
matching working memory tasks.

For the retinal data, on average $\pcf$ accounts for more than $75\%$ of pairwise correlations,
and $\pcs$ accounts for over $54\%$ of three-cell correlations. The interaction matrix of $\pcs$ contains information about the 
functional interaction features in the retinal circuitry. It seems that a retinal neuron can be paired with not only negatively strong couplings, but also
slightly positive couplings. Only a few pairs of
neurons have strong correlations with the global activity of the population. To describe the spike synchrony distribution,
$\pcs$ performs better than $\pcf$, nevertheless, both of them could not capture the trend of the tail (rare events related to
higher order interactions existing in the network). This is not surprising, because $\pcf$ and $\pcs$ are simplified descriptions of 
the original high dimensional neural activity, taking the trade-off between the computation complexity and the description goodness.

To extract the statistical structure embedded in the neural population, $\pcs$ improves significantly over $\pcf$, and has further additional
benefit of describing the third-order correlations observed in the data, as $\pcs$ could be used to construct triplet interactions among
neurons, although direct constructing all possible triplet interactions is extremely computationally difficult. 

Unlike the retinal
circuit, the cortical circuit yields a much smaller absolute value of the multi-information, implying that no significant
higher order correlations (interactions) were present in the neural circuit when the circuit was carrying out task-related information processing rather
than encoding well-structured stimuli (as in the retinal network). This also explains why a simplified description such as $\pcf$ and $\pcs$ is accurate enough to
capture the main features of the population activity, including the spike synchrony distribution. The inferred model on the 
cortical data reveals a different interaction map from that of the retinal circuit. In the cortical circuit, neurons form
the stripe-like structure in the interaction matrix, suggesting that these neurons may receive a large number of excitatory inputs~\cite{Fuji-2008}. These inputs may
come from different layers of cortex, and they can execute top-down or bottom-up information processing, thus modulate the global 
brain state in the target cortex during behavioral tasks.

Before summary of this work, I made some discussions about two relevant recent studies on population coupling (see {\em notes added}). Ref.~\cite{Mora-2016} modeled directly the joint probability distribution
of individual neural response and population rate (the number of neurons simultaneously active) by linear coupling and complete coupling models.
The linear coupling reproduces separately the distribution of individual neuronal state and the population rate distribution,
and their couplings, while $\pcf$ introduced in my
work reproduces mean firing rate and the correlation between individual neuronal state and the background population activity (except the neuron itself). Note that $\pcf$ does not model population rate distribution explicitly, which is hard to interpret in terms of
functional connectivity. The complete-coupling model 
reproduces the joint probability distributions between the response of each neuron and the population rate, 
from which it is hard to conclude that the high-order interactions responsible for high-order correlations can be interpreted and tested. However, $\pcs$ 
reproduces mean firing rate and
the correlation between neuron-pair activities and the background
population activity (except the neuron-pair itself), and thus explains high-order correlations by an energy model. Furthermore, in this sense, 
this work overcame a weakness pointed out in another independent later work of population coupling~\cite{Donnell-2016}, which fitted directly the population rate distribution and 
the firing probability for each neuron conditioned on the population rate, and analogously
the corresponding model parameters can not be readily interpretable in a biological setting. Due to intrinsic difference in model definitions, these two relevant works have nice properties of
studying tuning curves of individual neurons to the population rate, and sampling from the model to reproduce the population synchrony distribution.

In summary, I develop a theoretical model of population coupling and its advanced form, to relate the correlation profile in the 
neural collective activity to the basic circuit variables. The practical dimensional reduction method is tested on different types
of neural data, and specific features of neural circuit are revealed. This model aiming at describing high order correlations with a
low order representation, is expected to be useful for modeling big neural data. Note that the interaction matrices shown in Fig.~\ref{retinaPC2} and Fig.~\ref{mpcPC2}
are qualitatively robust to changes of the data size to only the first half (data not shown), verifying that the revealed features are not an artifact of overfitting. However, it still deserves further studies by
introducing regularization in the learning equation. It is also very interesting to incorporate more physiologically plausible parameters to explain how 
the collective spiking behavior arises from the microscopic interactions among the basic units. Another interesting study is to clarify 
the role of higher order correlations in decoding performances based on maximum likelihood principles~\cite{Isingdecoder-2012, Berry-2012}. 


\begin{acknowledgments}
I am grateful to Shigeyoshi Fujisawa and Michael J Berry for sharing me the cortical and retinal data, respectively. 
I also thank Hideaki Shimazaki and Taro Toyoizumi for stimulating discussions.
This work was supported by the program for Brain Mapping by Integrated Neurotechnologies 
for Disease Studies (Brain/MINDS) from Japan Agency for Medical Research and development, AMED.
\end{acknowledgments}

{\em note added.---}After I submitted this work to arXiv:1602.08299, I became aware of Ref.~\cite{Mora-2016} (arXiv:1606.08889), and later Ref.~\cite{Donnell-2016} (bioRxiv, 2016).
Discussions about these two recent relevant works are made in the last section of this paper.
\appendix
\section{Derivation of mean-field equations} \label{app:Deriv}
In this appendix, I give a simple derivation of mean-field equations given in Sec.~\ref{model}. More details can be obtained from Refs~\cite{Huang-2009pre,MM-2009}.
First, after removing an interaction $a$, one defines the cavity probability $P_{i\rightarrow a}(s_i)=\frac{e^{h_is_i}}{Z_i}\prod_{b\in\partial i\backslash a}\hat{P}_{b\rightarrow i}(s_i)$, where the product comes from
the physical meaning of the second kind of cavity probability, namely $\hat{P}_{b\rightarrow i}(s_i)$ defined as the cavity probability when only the connection from $b$ to $i$ is retained while other neighbors of $i$ are removed
(so-called cavity probability). $\hat{P}_{b\rightarrow i}(s_i)$ is thus formulated as $\sum_{s_j:j\in\partial b\backslash i}e^{\Gamma_b\prod_{j\in\partial b}s_j}\prod_{j\in\partial b\backslash i}P_{j\rightarrow b}(s_j)$. With these two probabilities,
it follows that the cavity magnetization $m_{i\rightarrow a}=\sum_{s_i}s_iP_{i\rightarrow a}(s_i)=\tanh\left(h_i+\sum_{b\in\partial i\backslash a}u_{b\rightarrow i}\right)$, where $u_{b\rightarrow i}$ is named cavity bias in physics~\cite{cavity-2001}. It is 
related to $\hat{m}_{b\rightarrow i}$ by $\hat{m}_{b\rightarrow i}=\tanh(u_{b\rightarrow i})=\sum_{s_i}s_i\hat{P}_{b\rightarrow i}(s_i)$.

More specifically, the cavity magnetization is derived as follows,
\begin{subequations}\label{magSM}
\begin{align}
\begin{split}
m_{i\rightarrow a}&=P_{i\rightarrow a}(+1)-P_{i\rightarrow a}(-1)=\frac{e^{h_i}\prod_{b\in\partial i\backslash a}\hat{P}_{b\rightarrow i}(+1)-e^{-h_i}\prod_{b\in\partial i\backslash a}\hat{P}_{b\rightarrow i}(-1)}
{e^{h_i}\prod_{b\in\partial i\backslash a}\hat{P}_{b\rightarrow i}(+1)+e^{-h_i}\prod_{b\in\partial i\backslash a}\hat{P}_{b\rightarrow i}(-1)}\\
&=\frac{e^{h_i}-e^{-h_i-2\sum_{b\in\partial i\backslash a}u_{b\rightarrow i}}}{e^{h_i}+e^{-h_i-2\sum_{b\in\partial i\backslash a}u_{b\rightarrow i}}}\\
&=\tanh\left(h_i+\sum_{b\in\partial i\backslash a}u_{b\rightarrow i}\right),
\end{split}
\end{align}
\end{subequations}
where I have used the definition $u_{b\rightarrow i}\equiv\frac{1}{2}\ln\frac{\hat{P}_{b\rightarrow i}(+1)}{\hat{P}_{b\rightarrow i}(-1)}$. Next, I show how to derive the cavity bias. First, for 
pairwise interaction,
\begin{subequations}\label{biasSM}
\begin{align}
\begin{split}
u_{b\rightarrow i}&=\frac{1}{2}\ln\frac{e^{\Gamma_b}P_{j\rightarrow b}(+1)+e^{-\Gamma_b}P_{j\rightarrow b}(-1)}{e^{-\Gamma_b}P_{j\rightarrow b}(+1)+e^{\Gamma_b}P_{j\rightarrow b}(-1)}\\
&=\frac{1}{2}\ln\frac{1+\tanh\Gamma_b m_{j\rightarrow b}}{1-\tanh\Gamma_b m_{j\rightarrow b}}\\
&=\tanh^{-1}\left(\tanh\Gamma_b m_{j\rightarrow b}\right),
\end{split}
\end{align}
\end{subequations}
where I have used the parameterization $P_{j\rightarrow b}(s_j)=\frac{1+s_jm_{j\rightarrow b}}{2}$, and the mathematical identity $e^{2z}=\frac{1+\tanh z}{1-\tanh z}$.
Similarly, for triplet interaction,
\begin{subequations}\label{biasSM02}
\begin{align}
\begin{split}
u_{b\rightarrow i}&=\frac{1}{2}\ln\frac{e^{\Gamma_b}[P_{j\rightarrow b}(+1)P_{k\rightarrow b}(+1)+P_{j\rightarrow b}(-1)P_{k\rightarrow b}(-1)]+e^{-\Gamma_b}[P_{j\rightarrow b}(+1)P_{k\rightarrow b}(-1)+P_{j\rightarrow b}(-1)P_{k\rightarrow b}(+1)]}
{e^{-\Gamma_b}[P_{j\rightarrow b}(+1)P_{k\rightarrow b}(+1)+P_{j\rightarrow b}(-1)P_{k\rightarrow b}(-1)]+e^{\Gamma_b}[P_{j\rightarrow b}(+1)P_{k\rightarrow b}(-1)+P_{j\rightarrow b}(-1)P_{k\rightarrow b}(+1)]}\\
&=\frac{1}{2}\ln\frac{1+e^{-2\Gamma_b}\frac{1-m_{j\rightarrow b}m_{k\rightarrow b}}{1+m_{j\rightarrow b}m_{k\rightarrow b}}}{e^{-2\Gamma_b}+\frac{1-m_{j\rightarrow b}m_{k\rightarrow b}}{1+m_{j\rightarrow b}m_{k\rightarrow b}}}\\
&=\tanh^{-1}\left(\tanh\Gamma_b m_{j\rightarrow b}m_{k\rightarrow b}\right).
\end{split}
\end{align}
\end{subequations}
These results are written in a compact form as Eq.~(\ref{bp}) in the main text.

Similarly, the single neuron magnetization $m_i$ is obtained via $m_i=\frac{1}{Z_i}\sum_{s_i}s_ie^{h_is_i}\prod_{b\in\partial i}\hat{P}_{b\rightarrow i}(s_i)$, where $Z_i$ is a normalization constant,
and the multi-neuron correlation $C_a=\frac{1}{Z_a}\sum_{s_i:i\in\partial a}
\prod_{i\in\partial a}s_ie^{\Gamma_a\prod_{i\in\partial a}s_i}\prod_{i\in\partial a}P_{i\rightarrow a}(s_i)$, where $Z_a$ is a normalization constant. Note that $Z_i$ and $Z_a$ are also related to the free
energy contribution of single neuron and neuronal interaction~\cite{Huang-2009pre}, respectively. The full (non-cavity) magnetization can be derived in a similar manner to Eq.~(\ref{magSM}), as $m_i=\tanh\left(h_i+\sum_{b\in\partial i}u_{b\rightarrow i}\right)$. In detail, the two-point correlation $C_a$ ($|\partial a|=2$) is computed as follows,
\begin{subequations}\label{twocorre}
\begin{align}
\begin{split}
C_a&\equiv\left<s_is_j\right>=\frac{\sum_{s_i,s_j}s_is_je^{\Gamma_a s_is_j}P_{i\rightarrow a}(s_i)P_{j\rightarrow a}(s_j)}{\sum_{s_i,s_j}e^{\Gamma_a s_is_j}P_{i\rightarrow a}(s_i)P_{j\rightarrow a}(s_j)}\\
&=\frac{e^{2\Gamma_a}-\frac{1-m_{j\rightarrow a}m_{i\rightarrow a}}{1+m_{j\rightarrow a}m_{i\rightarrow a}}}{e^{2\Gamma_a}+\frac{1-m_{j\rightarrow a}m_{i\rightarrow a}}{1+m_{j\rightarrow a}m_{i\rightarrow a}}}\\
&=\frac{\tanh\Gamma_a+m_{i\rightarrow a}m_{j\rightarrow a}}{1+\tanh\Gamma_a m_{i\rightarrow a}m_{j\rightarrow a}}.
\end{split}
\end{align}
\end{subequations}
Analogously, three-point correlation ($|\partial a|=3$) can be evaluated as 
\begin{subequations}\label{threecorre}
\begin{align}
\begin{split}
C_a&\equiv\left<s_is_js_k\right>=\frac{\sum_{s_i,s_j,s_k}s_is_js_ke^{\Gamma_a s_is_js_k}P_{i\rightarrow a}(s_i)P_{j\rightarrow a}(s_j)P_{k\rightarrow a}(s_k)}{\sum_{s_i,s_j,s_k}e^{\Gamma_a s_is_js_k}P_{i\rightarrow a}(s_i)P_{j\rightarrow a}(s_j)P_{k\rightarrow a}(s_k)}\\
&=\frac{e^{2\Gamma_a}-\frac{1-m_{j\rightarrow a}m_{i\rightarrow a}m_{k\rightarrow a}}{1+m_{j\rightarrow a}m_{i\rightarrow a}m_{k\rightarrow a}}}{e^{2\Gamma_a}+\frac{1-m_{j\rightarrow a}m_{i\rightarrow a}m_{k\rightarrow a}}{1+m_{j\rightarrow a}m_{i\rightarrow a}m_{k\rightarrow a}}}\\
&=\frac{\tanh\Gamma_a+m_{i\rightarrow a}m_{j\rightarrow a}m_{k\rightarrow a}}{1+\tanh\Gamma_a m_{i\rightarrow a}m_{j\rightarrow a}m_{k\rightarrow a}}.
\end{split}
\end{align}
\end{subequations}
These results are written in a compact form as Eq.~(\ref{magcorre}b) in the main text.

Finally, the partition function $Z_i$ for pairwise interaction is computed as follows,
\begin{subequations}\label{Zipair}
\begin{align}
\begin{split}
Z_i&=e^{h_i}\prod_{b\in\partial i}\hat{P}_{b\rightarrow i}(+1)+e^{-h_i}\prod_{b\in\partial i}\hat{P}_{b\rightarrow i}(-1)\\
&=e^{h_i}\prod_{b\in\partial i}\cosh\Gamma_b(1+\tanh\Gamma_b m_{j\rightarrow b})+e^{-h_i}\prod_{b\in\partial i}\cosh\Gamma_b(1-\tanh\Gamma_b m_{j\rightarrow b}),
\end{split}
\end{align}
\end{subequations}
where I have used $\hat{P}_{b\rightarrow i}(s_i)=e^{\Gamma_b s_i}P_{j\rightarrow b}(+1)+e^{-\Gamma_b s_i}P_{j\rightarrow b}(-1)$ and the magnetization parameterization of $P_{j\rightarrow b}$.
For triplet interaction, $\hat{P}_{b\rightarrow i}(s_i)=e^{\Gamma_b s_i}[P_{j\rightarrow b}(+1)P_{k\rightarrow b}(+1)+P_{j\rightarrow b}(-1)P_{k\rightarrow b}(-1)]+e^{-\Gamma_b s_i}[P_{j\rightarrow b}(+1)P_{k\rightarrow b}(-1)+P_{j\rightarrow b}(-1)P_{k\rightarrow b}(+1)]$, 
and the corresponding $Z_i=e^{h_i}\prod_{b\in\partial i}\cosh\Gamma_b(1+\tanh\Gamma_b m_{j\rightarrow b}m_{k\rightarrow b})+e^{-h_i}\prod_{b\in\partial i}\cosh\Gamma_b(1-\tanh\Gamma_b m_{j\rightarrow b}m_{k\rightarrow b})$.
Following the same line, the partition function $Z_a$ is evaluated for pairwise interaction as
\begin{subequations}\label{Zapair}
\begin{align}
\begin{split}
Z_a&=\sum_{s_i,s_j}e^{\Gamma_a s_is_j}P_{i\rightarrow a}(s_i)P_{j\rightarrow a}(s_j)\\
&=\cosh\Gamma_a[1+\tanh\Gamma_a m_{i\rightarrow a}m_{j\rightarrow a}].
\end{split}
\end{align}
\end{subequations}
Similarly, the partition function $Z_a$ for triplet interaction can be computed as
\begin{subequations}\label{Zatri}
\begin{align}
\begin{split}
Z_a&=\sum_{s_i,s_j,s_k}e^{\Gamma_a s_is_js_k}P_{i\rightarrow a}(s_i)P_{j\rightarrow a}(s_j)P_{k\rightarrow a}(s_k)\\
&=\cosh\Gamma_a[1+\tanh\Gamma_a m_{i\rightarrow a}m_{j\rightarrow a}m_{k\rightarrow a}],
\end{split}
\end{align}
\end{subequations}
which is exactly the compact equation given in the main text.




\end{document}